\documentclass[prd,aps,showpacs,tightenlines,twocolumn]{revtex4}
\usepackage{graphicx}

\begin{document}

\title{MSSM flat direction as a curvaton}
\author{Kari Enqvist$^{a,b}$, Asko Jokinen$^b$, Shinta Kasuya$^b$,
        and Anupam Mazumdar$^c$}

\affiliation{
$^a$ Department of Physical Sciences,
     P. O. Box 64, FIN-00014, University of Helsinki, Finland.\\
$^b$ Helsinki Institute of Physics, P. O. Box 64,
     FIN-00014, University of Helsinki, Finland.\\
$^c$ Physics Department, McGill University,
     3600-University Road, Montreal, H3A 2T8, Canada.}

\date{March 19, 2003}

\begin{abstract}

We study in detail the possibility that the flat directions of the
Minimal Supersymmetric Standard Model (MSSM) could act as a curvaton
and generate the observed adiabatic density perturbations. For that
the flat direction energy density has to dominate the Universe at the
time when it decays. We point out that this is not possible if the inflaton
decays into MSSM degrees of freedom. If the inflaton is completely in
the hidden sector, its decay products do not couple to the flat
direction, and the flat direction curvaton can dominate the energy
density. This requires the absence of a Hubble-induced mass for the
curvaton, e.g. by virtue of  the Heisenberg symmetry. In the case of
hidden radiation, $n=9$ is the only admissible direction; for other
hidden equations of state, directions with lower $n$ may also
dominate. We show that the MSSM curvaton is further constrained
severely by the damping of the fluctuations, and as an example,
demonstrate that in no-scale supergravity it would fragment into $Q$
balls rather than decay. Damping of fluctuations can be avoided by an
initial condition, which for the $n=9$ direction would require an
initial curvaton amplitude of $\sim 10^{-2}M_p$, thereby providing
a working example of the MSSM flat direction curvaton.
\end{abstract}

\pacs{98.80.Cq
\hspace{37mm} HIP-2003-13/TH, hep-ph/0303165}

\maketitle

\section{Introduction}

The Minimal Supersymmetric Standard Model (MSSM) is well known to have
flat directions, made up of squarks and sleptons, along which
the scalar potential vanishes above the soft supersymmetry breaking
scale~$\sim 1$~TeV \cite{Dine:1995kz,Gherghetta:1995dv}. The MSSM flat
directions have important cosmological consequences for the early
Universe and may seed Affleck-Dine baryo/leptogenesis
\cite{Affleck:1984fy,Dine:1995uk,Dine:1995kz}, give rise to non-thermal
generation of supersymmetric dark matter
\cite{Kusenko:1997si,Enqvist:1998en} or $B$-ball baryogenesis below
the electroweak scale~\cite{Enqvist:1997si}, and may also act as a
source for isocurvature density perturbations
\cite{Enqvist:1998pf,Enqvist:1999mv,Kawasaki:2001in}
(for a review, see \cite{Enqvist:2003gh}).

Inflation wipes out all the inhomogeneities along a given flat
direction, leaving only the zero mode condensate.  However, during
inflation quantum fluctuations along the flat directions impart
isocurvature density perturbations on the condensate
\cite{Enqvist:1998pf}. The isocurvature fluctuations can later be
converted into adiabatic perturbations at the time when the flat
direction decays into the radiation of the MSSM degrees of freedom
\cite{Enqvist:2002rf}, provided the flat direction dominates the
energy density of the Universe at the time of the decay. Obviously,
during inflation the flat direction should be subdominant and its
mass should be smaller than the Hubble parameter. This is an example of
the so-called curvaton scenario, which in its present incarnation was
first discussed in the context of pre-big bang \cite{Enqvist:2001zp}
and then applied to ordinary inflation \cite{Lyth:2001nq}. In many
early papers \cite{Lyth:2001nq,Others} the curvaton potential was
simply taken as a quadratic potential $V=m^2\phi^2$.

For the MSSM flat direction curvaton, the potential is determined by
the supersymmetry breaking but it is usually dominated by nonrenormalizable
operators at large amplitudes. There are three conditions such a potential
should satisfy. First, the energy density of the flat direction must
contribute negligibly during inflation but should dominate the Universe
at later time when the flat direction decays into radiation. Second, the
flat direction field should stay in the right place to yield the right
amount of (isocurvature) fluctuations. Third, the fluctuations produced
during inflation must not die out during the whole process. In
\cite{Enqvist:2002rf}, we have studied the conditions for the later
energy domination, and partly the amplitude of the fluctuations during
inflation. Studying the consequences of all these constraints
comprehensively is the main purpose of this article.

The structure of the paper is as follows. In Sect. II we discuss
general constraints which the flat directions need to obey in order
to be viable curvatons. In Sect.III, we follow the dynamics of the flat
direction, and study which directions may act as a curvaton. The
behavior of the fluctuations is considered in Sect.IV. In Sect.V, we
study the MSSM flat direction in no-scale supergravity, and show that
$Q$-ball formation is inevitable. Section VI is devoted to our
conclusions.


\section{Flat direction and inflation}

The degeneracy of the effective potential of the MSSM flat direction
is lifted by supersymmetry (SUSY) breaking effects and some nonrenormalizable
operators. In general, we can thus write the potential as
\begin{eqnarray}
    V(\phi) & = &\frac{1}{2}m_{\phi}^2\phi^2 + V_{NR}\,, \\
    V_{NR} & = & \frac{\lambda^2 \phi^{2(n-1)}}{2^{n-1}M^{2(n-3)}}\,,
\end{eqnarray}
where $m_{\phi}\sim$~TeV is the soft SUSY breaking mass, and the flat
direction condensate is $\Phi=\phi e^{i \theta}/\sqrt{2}$. $M$ is the
cut-off scale for the low energy effective theory, usually taken to be
the Planck scale $M_p \simeq 2.4\times 10^{18}$ GeV; $\lambda$ is a
coupling constant; and $n=4,\dots , 9$ is the dimension of the
nonrenormalizable operator lifting the flat direction, the value of
which depends on which particular flat direction one is discussing
(for details, see \cite{Enqvist:2003gh}).

In supergravity (SUGRA) theories, the flat direction often acquires the
mass of order $H$ because of the SUSY breaking effect due to the finite
vacuum energy during inflation \cite{Dine:1995uk}. If so, the
fluctuation amplitude along the flat direction dies out completely
during inflation. We thus demand that the inflation model is such
that mass term as large as $H$ is not induced. One example is the
SUSY D-term inflation, which  during inflation leads to a vanishing
Hubble-induced mass term for the flat directions \cite{Kolda:1998kc}.
Another example is models obtained from SUGRA theories with a
Heisenberg symmetry on the K\"ahler manifold \cite{Gaillard:1995az}.
These give rise to a K\"ahler potential of the form
\begin{equation}
    \label{Kpot}
    G=f(\eta)+\ln\left| W(\phi_i)\right|^2+g(y_a)\,,
\end{equation}
with $\eta=z+z^*-\phi_i^*\phi_i$, where $z$ is the Polonyi field, and
$\phi_i$ and $y_a$ are respectively the observable and hidden
fields. The latter are defined as the ones that have only
Planck-suppressed couplings to the observable sector. In this case
there is no mass term in the tree-level potential for the flat
direction. No-scale models \cite{Lahanas:1986uc}, for which
$f(\eta)=-3\ln\eta$, are a particular realization of the K\"ahler
manifold Heisenberg symmetry. However, even with a Heisenberg symmetry
there will be radiatively induced mass squared which is small and
negative with $m_{\phi,eff}^2 \sim -10^{-2} H^2$
\cite{Gaillard:1995az}. Such a small mass term has only negligible
damping effect on the fluctuation amplitude.

In what follows we simply assume that during inflation the flat
direction does not get any appreciable Hubble-induced mass, e.g. by
virtue of the Heisenberg symmetry, or by some other reasons.

In order for the curvaton scenario to work, fluctuations of the
inflaton should not contribute significantly to the adiabatic
density perturbations, so that the Hubble parameter during inflation
is $H_* \sim \rho_{inf}^{1/2}M_p< 10^{14}$ GeV. (Needless to say, the
energy density of the flat direction should be negligible compared to
that of the inflaton, $\rho_{\phi} \ll \rho_{inf}$). Then the
isocurvature fluctuation of the flat direction is
$\delta\phi \sim H_*/2\pi$. If $\delta\phi/\phi_* \sim H_*/\phi_*
\sim 10^{-5}$, where $\phi_*$ is the amplitude during inflation,
obtained from $V''(\phi_*) \sim H_*^2$, the right amount of
density perturbation can be generated provided  there is no later
damping. A simple analysis shows that during inflation the flat
direction field condensate is slow-rolling in the nonrenormalizable
potential $V_{NR}$. Thus, the Hubble parameter and the amplitude of
the field can respectively be estimated as,
\begin{eqnarray}
    H_* & \sim & \lambda^{-\frac{1}{n-3}}\delta^{n-2 \over n-3}M_p\,,\\
    \phi_* & \sim & \lambda^{-\frac{1}{n-3}}\delta^{1 \over n-3}M_p\,,
\end{eqnarray}
where $\delta\equiv \delta\phi/\phi_* \sim H_*/\phi_*$.

After inflation, the inflaton ultimately decays into relativistic
degrees of freedom. A priori, there are two possibilities. Because the
inflaton should give rise to all the observable baryons,
conventionally one usually assumes that the inflaton must decay into
particles of the observable sector. The other possibility, discussed
within the context of the MSSM curvaton scenario \cite{Enqvist:2002rf},
is to assume that the inflaton decays into the hidden sector and that
the baryons originate solely from the flat direction curvaton decay.

If the inflaton decay products consist of (MS)SM particles, one should
consider the behavior of the flat direction in a thermal background
which interacts with the condensate field. It has been argued by
Postma \cite{Postma:2002et} that the flat direction condensate decays
by thermal scattering before its domination. However, in her analysis
the thermal decay rate was taken to be $\sim f^4T^2/m$, whereas in a
thermal environment $m$ should be replaced by $fT$ ($f$ is here some
coupling). Nevertheless, the conclusion remains essentially the same,
as we now argue. The energy density (amplitude) of the flat direction
field in $V \sim T^2\phi^2$ behaves as
$\rho_{\phi} \propto a^{-27/8}(\phi \propto a^{-21/16})$ during the
inflaton-oscillation dominated Universe, while
$\rho_{\phi} \propto a^{-4} (\phi \propto a^{-1})$ during radiation
domination. In either case, its energy density decreases not slower
than that of radiation, and the amplitude becomes so small that the
flat direction condensate cannot dominate the energy density after the
zero temperature part $m_{\phi}^2\phi^2$ becomes important.

These difficulties can be avoided if one takes the inflaton sector to
be completely decoupled from the observable one \cite{Enqvist:2002rf}.
Indeed, there is not a single realistic particle physics model which
would embody the inflaton into the family of the observable fields. In
almost all the models the inflaton is a gauge singlet which largely
lives in the isolated inflaton sector as if it were part of a hidden
world. The coupling of such a singlet to the SM degrees of freedom is
usually set by hand. Under such circumstances, perhaps a hidden sector
inflaton would be a logical conclusion. Such an inflaton would decay
into (light) particles in the hidden sector, but the hidden thermal
background would not interact with the flat direction condensate.
Note that the reheating of the observable degrees of freedom in the
Universe takes place due to the decay of the MSSM flat direction into
the MS(SM) degrees of freedom. The curvaton mechanism works
successfully if this temperature is high enough. We will come back to
this point later.

\section{Late domination of the flat direction energy density}

For a successful MSSM curvaton scenario, the energy density of the
flat direction condensate should dominate the Universe at the time of
its decay. The condensate starts to oscillate when $H \sim m_{\phi}$,
and the amplitude at that time is
\begin{equation}
    \label{osc}
    \phi_{osc} \sim \left(\frac{m_{\phi}M^{n-3}}{\lambda}
    \right)^{1 \over n-2}\,.
\end{equation}
If the reheating by the hidden inflaton occurs earlier, the Universe
is dominated by hidden radiation at this time. Since
the energy density of the condensate is $\propto a^{-3}\propto H^{-3/2}$,
we find that
\begin{eqnarray}
    \rho_{\phi}\big|_{EQ} & \sim & m_{\phi}^2\phi_{osc}^2
    \left(\frac{H_{EQ}}{H_{osc}}\right)^{3/2} \nonumber \\
    & \sim & \left(m_{\phi} H_{EQ}^3\right)^{1/2}
    \left(\frac{m_{\phi}M^{n-3}}{\lambda}\right)^{2 \over n-2}\,,
\end{eqnarray}
at the time it equals the hidden radiation density
$\rho_h \sim H_{EQ}^2 M_p^2$. Thus the Hubble parameter at the
equality time is
\begin{equation}
    H_{EQ} \sim m_{\phi}\left(\frac{m_{\phi}M^{n-3}}
      {\lambda M_p^{n-2}}\right)^{4 \over n-2}\,.
\end{equation}
In order for the flat direction condensate to have a chance to
dominate the Universe, $H_{EQ}$ should be larger than $H$ at the time
the curvaton condensate decays. If the decay rate is written as
$\Gamma_{\phi} \sim f^2 m_{\phi}$, where $f$ is some Yukawa or gauge
coupling, we have a constraint on the coupling constant which reads
\begin{equation}
    \label{ef}
    f < \lambda^{-\frac{2}{n-2}}
    \left(\frac{m_{\phi}}{M_p}\right)^{\frac{2}{n-2}}
    \left(\frac{M}{M_p}\right)^{\frac{2(n-3)}{n-2}}\,.
\end{equation}

In the opposite case, when the reheating in the hidden sector
occurs after the oscillation of the flat direction started,
i.e., $H_{osc}>H_{RH}$, the energy density evolves as
\begin{eqnarray}
    \rho_{\phi}\big|_{EQ} & \sim & \rho_{\phi}\big|_{osc}
    \left(\frac{H_{RH}}{H_{osc}}\right)^2
    \left(\frac{H_{EQ}}{H_{RH}}\right)^{3/2}\,, \nonumber \\
    & \sim & \left(\frac{m_{\phi}M^{n-3}}{\lambda}\right)^{2\over n-2}
    \left(\frac{T_{RH}^2}{M_p}\right)^{1/2} H_{EQ}^{3/2}\,,
\end{eqnarray}
so that taking into account $H_{osc}>H_{RH}$ we actually get the same
constraint as in Eq. (\ref{ef}).

Even for a small coupling of the order of the electron Yukawa coupling
such as $f \sim 10^{-6}$, all $n \leq 6$ cases fail to satisfy the
condition Eq. (\ref{ef}), whereas $n=7$ is marginal when
$\lambda \sim 1$.

\begin{figure}
\includegraphics[width=80mm]{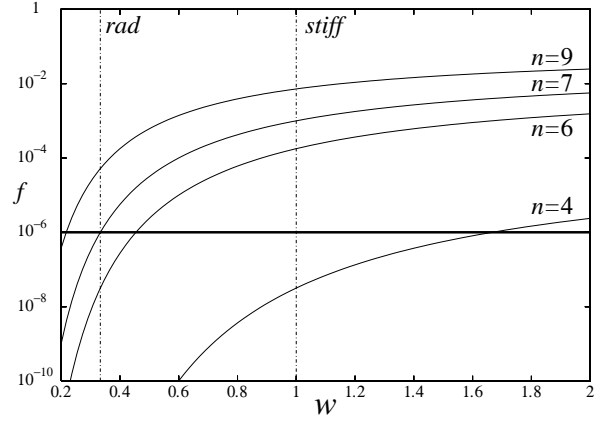}
\caption{\label{hid}
Constraint on the coupling for $n=4,6,7$, and 9. The allowed region is
below these lines and above $f=10^{-6}$.}
\end{figure}

The condition Eq. (\ref{ef}) depends on the equation of state of the
inflaton decay products in the hidden sector.  Let us therefore write
the equation of state as $p_{h}=w\rho_{h}$, and assume that the hidden
energy has already dominated the Universe when the oscillations along
the flat direction begin. Then we obtain the ratio of the energy densities
\begin{eqnarray}
    \frac{\rho_{\phi}}{\rho_h}
    & \sim & \left.\frac{\rho_{\phi}}{\rho_h}\right|_{osc}
    \left(\frac{H}{H_{osc}}\right)^{-\frac{2w}{1+w}}\,, \nonumber \\
    & \sim & \left(\frac{m_{\phi}}{\lambda M_p}\right)^{2 \over n-2}
    \left(\frac{H}{H_{osc}}\right)^{-\frac{2w}{1+w}}\,.
\end{eqnarray}
This ratio becomes unity when $H \sim H_{EQ}$. Imposing
$H_{EQ}>\Gamma_{\phi} \sim f^2 m_{\phi}$, we obtain
\begin{equation}
    f < \left[ \left(\frac{m_{\phi}}{\lambda M_p}
      \right)^{1 \over n-2}\right]^{\frac{1+w}{2w}}\,.
\end{equation}
Notice that this is the same as Eq.(\ref{ef}) for $w=1/3$.
We show the constraints for $n=4,6,7,$ and 9 in Fig.~\ref{hid}. As
discussed in \cite{Enqvist:2002rf}, $n=9$ direction is essentially the
only viable option for the hidden radiation case, but even $n=6$ directions
can be acceptable if the hidden sector fluid has a stiff equation of
state ($w=1$). Notice that $n=4$ directions can never dominate the
Universe at any point and are thus completely ruled out as a curvaton
candidate.

\section{Evolution of perturbations}

So far we have assumed that the isocurvature perturbation created
during inflation does not evolve. This is strictly true in
$m_{\phi}^2 \phi^2$ potential, since both the homogeneous and the (linear)
perturbation parts obey the same equations of motion. Here we shall see
whether this assumption holds in more general cases.

\subsection{Slow rolling in nonrenormalizable potential}

For the MSSM curvaton scenario to work, the flat direction must have
a vanishing (or more precisely, negligible) mass during inflation. In
such a case, the field will be slow-rolling in the nonrenormalizable
potential $V_{NR} \sim \lambda^2\phi^{2(n-1)}/M^{2(n-3)}$. In
addition, we assume here that there is no Hubble-induced mass term
even after inflation, so that the field will continue slow-rolling in
$V_{NR}$ down to the amplitude $\phi_{osc}$, which is determined by
$V_{NR}(\phi_{osc}) \sim m_{\phi}^2\phi_{osc}^2$. In general, the
equations of motion for the homogeneous and fluctuation parts are
written respectively as
\begin{eqnarray}
    & & \ddot{\phi}+3H\dot{\phi}+V'(\phi) = 0\,, \\
    & & \delta\ddot{\phi}_k+3H\delta\dot{\phi}_k
    + \frac{k^2}{a^2}\delta\phi_k+V''(\phi) \delta\phi_k = 0\,,
\end{eqnarray}
where the prime denotes the derivative with respect to $\phi$. Since
we are interested only in the super horizon mode ($k\rightarrow 0$),
using the slow roll approximation we have
\begin{eqnarray}
    \label{sr-homo}
    & & 3H\dot{\phi}+V'(\phi) = 0\,, \\
    \label{sr-fl}
    & & 3H\delta\dot{\phi}+V''(\phi) \delta\phi = 0\,.
\end{eqnarray}
Hereafter we omit the subscript $k$, understanding that $\delta\phi$
is for the super horizon mode. Then it is easy to obtain the evolution
of the ratio of the fluctuation and the homogeneous mode in a
$V_{NR} \propto \phi^{2(n-1)}$ potential. The result is:
\begin{equation}
    \frac{\delta\phi}{\phi} \sim
    \left(\frac{\delta\phi}{\phi}\right)_i
    \left(\frac{\phi}{\phi_i}\right)^{2(n-2)}\,,
\end{equation}
where $i$ denotes the initial values.

During inflation the homogeneous field obeys Eq. (\ref{sr-homo}),
which can be easily integrated to yield
\begin{equation}
    \frac{\phi}{\phi_i} \simeq \left( 1 +
      \frac{1}{3(2n-3)} \ \frac{V''(\phi_i)}{H^2}\Delta N
      \right)^{-\frac{1}{2(n-2)}}\,,
\end{equation}
where $\Delta N$ is the number of e-folds. Since we are concerned with
the slow-roll regime, it is reasonable to require
$V''(\phi_i)/H^2 \lesssim 1$. Hence we have $\phi/\phi_i \approx 0.95$
for the last 50 e-folds in $n=9$ case, for example. This implies that
the amplitude of the fluctuation relative to its homogeneous part
decreases only by factor $\simeq 2$. Hence during this stage there is
essentially no damping. Notice that the slower the condensate field
rolls during the last 50 e-folds, the less damping there is.

After inflation the curvaton condensate slow-rolls (albeit marginally),
i.e., $V''(\phi) \sim H^2$, and we can still use the slow-roll approximation
equations (\ref{sr-homo}) and (\ref{sr-fl}). During this stage, the field
amplitude is given by $\phi \sim (H M^{n-3}/\lambda)^{1/(n-2)}$, while the
Hubble parameter changes from $H_*$ to $m_{\phi}$. As a consequence, there
is a huge damping given by
\begin{equation}
    \frac{\left(\frac{\delta\phi}{\phi}\right)_{osc}}
    {\left(\frac{\delta\phi}{\phi}\right)_*} \sim
    \left(\frac{m_{\phi}}{H_*}\right)^2
    \sim \lambda^{2 \over n-3} \delta^{-\frac{2(n-2)}{n-3}}
    \left(\frac{m_{\phi}}{M}\right)^2\,,
\end{equation}
where we have used $H_* \sim \lambda^{-1/(n-3)}\delta^{(n-2)/(n-3)}M$
in the last equality, and $\delta \simeq H_*/\phi_*$ is the
fluctuation during inflation. Even for $n=4$, the damping factor is
$10^{-10}$ for $M=M_p$ and $\delta \sim 10^{-5}$. The situation is
still worse for the  directions with larger $n$. Hence the primordial
fluctuations of the MSSM flat direction curvaton appear to be
effectively wiped out. However, before drawing any definite
conclusions, one should also consider the effects of Hubble-induced
mass terms which can appear after inflation. This is the case, for
example, in D-term inflation.

\subsection{Positive Hubble-induced mass term}
The Hubble-induced effective potential can be  written as
\begin{equation}
    V_H = \frac{1}{2}c_H H^2\phi^2\,.
\end{equation}
The sign of the coefficient $c_H$ is usually determined by higher
order non minimal K\"ahler potential, so there are equal possibilities
for positive and negative mass term. Let us first consider the
positive case. When the Hubble-induced effective potential dominates,
the equations of motion for the homogeneous and the fluctuation mode
have the same form:
\begin{equation}
    \ddot{\psi}+3H\dot{\psi}+c_H H^2 \psi = 0\,,
\end{equation}
where $\psi=\phi$ or $\delta\phi$. From the viewpoint of the evolution
of the relative amplitude of the fluctuations, there is no damping.
However, the amplitude itself diminishes considerably. If $c_H>9/16$,
the decrease is $\phi\propto H^{1/2}$. At the onset of curvaton
oscillations, when $H\sim m_{\phi}$, the amplitude of the curvaton is
then
\begin{equation}
    \phi_{osc} \sim \lambda^{-\frac{1}{2(n-3)}}
    \delta^{-\frac{n-4}{2(n-3)}} \left(m_{\phi} M\right)^{\frac{1}{2}}\,.
\end{equation}
Since $\delta \gtrsim 10^{-5}$, a maximum is achieved for the largest
$n$. For $n=9$, $\phi_{osc} \sim 10^2 (m_{\phi}M)^{1/2}$, which is
just $10^2$ times larger than in the case of $n=4$, and $\sim 10^2$
times smaller than in the slow roll  $n=6$ case discussed in the
previous subsection. (Notice that the amplitude has the same behavior
as in the slow-roll case for $n=4$.) Thus, at the time of its decay,
the energy density of the flat direction condensate cannot dominate
the Universe.

\subsection{Negative Hubble-induced mass term}
In this case, the homogeneous field is trapped in the instantaneous
minimum $\phi_m \sim (H M^{n-3}/\lambda)^{1/(n-2)}\propto
H^{1/(n-2)}$. Then, up to a numerical factor, the equation of motion
for the fluctuation is identical to the positive Hubble-induced mass
case. Hence the amplitude of the fluctuation decreases as
$\delta\phi \propto H^{1/2}$. Therefore, the ratio of the amplitudes
of the the homogeneous and the fluctuation modes is given by
\begin{eqnarray}
    \left.\frac{\delta\phi}{\phi}\right|_{osc} & \sim &
    \left.\frac{\delta\phi}{\phi}\right|_*
    \left(\frac{m_{\phi}}{H_*}\right)^{\frac{n-4}{2(n-2)}}\,,
    \nonumber \\
    & \sim & \lambda^{\frac{n-4}{2(n-2)(n-3)}}
    \delta^{-\frac{n-4}{2(n-3)}}
    \left(\frac{m_{\phi}}{M}\right)^{\frac{n-4}{2(n-2)}}\,.
\end{eqnarray}
Since the field is released from the trap when $H \sim m_{\phi}$, the
subsequent energy domination condition is the same as discussed in the
previous section (See Fig.~\ref{hid}). Thus, the only viable direction
is $n=9$ (and  highly marginally $n=7$). We know that  for the
curvaton scenario $H_* < 10^{14}$ GeV so that the fluctuations of the
flat direction condensate during inflation cannot be too large:
$\delta < 3 (6) \times 10^{-4}$ for $n=9 (7)$. Thus, the conclusion is
that ratio of the fluctuation to the homogeneous mode is much less
than $10^{-5}$ at the onset of the flat direction condensate
oscillations.

One may wonder whether the damping effect becomes any milder for
$|c_H|<9/16$. Such a situation may be realized in the context of the
no-scale SUGRA. However, a small Hubble-induced mass term  reduces to
the case in subsection A, where the field slow-rolls in a
nonrenormalizable potential. Thus, the amplitude of the fluctuations
will be wiped out.

\subsection{The way out}

As seen above, the energy nondomination and/or the damping  of
the fluctuation amplitude usually kill the MSSM flat direction
curvaton scenario. Damping arises because the curvaton has to slide
from the slope of the non-renormalizable potential down to value at
which oscillations commence, a process which takes place slowly and is
thus associated with a considerable redshift.

We have found that there are essentially two ways to avoid all these
problems. One is that the coupling $\lambda$ of the nonrenormalizable
term is small enough so that the potential is effectively of the form
$m_{\phi}^2\phi^2$. In this case, during inflation  the amplitude of
the flat direction is $\phi_* \sim M_p$ with $H_* \sim 10^{13}$ GeV.
The other possibility is that the field amplitude at the end of
inflation happens to be of the same order as the amplitude
$\phi_{osc}$. Such a situation may be realized by an extremely long
period of inflation, or simply by chance. For the $n=9$ direction,
$\phi_{osc} \sim 10^{16}$ GeV, which is only an order of magnitude
less than the ``natural'' value for  $\phi_*$. It is conceivable that
such a low value of $\phi_*$ could be given e.g. by some chaotic
initial conditions. Hence we may conclude that hidden inflation with a
MSSM curvaton can indeed provide the correct adiabatic density
perturbations, although with some difficulty.

\section{Fragmentation of the flat direction}

One should also consider the dynamics of the curvaton after its
oscillations begin. So far, we have not taken into account the running
of the mass of the flat direction. In general, in the gravity mediated
SUSY breaking case the mass term in the effective potential can be
written as \cite{Enqvist:1998en}
\begin{equation}
  V(\phi) \simeq m_{\phi}^2\left[1
    +K\log\left(\frac{\phi^2}{M^2}\right)\right]\phi^2\,,
\end{equation}
where $K$ is a coefficient obtained from one-loop corrections. MSSM
curvaton dynamics is complicated by the fact that when $K$ is
negative, the flat direction condensate naturally fragments into $Q$
balls soon after it starts the oscillations
\cite{Enqvist:1998en,Kasuya:1999wu}. Then the isocurvature
fluctuations of the condensate remain trapped in the $Q$ balls, and
will only be released through the decay of the $Q$ balls. They will be
converted into adiabatic perturbations only if the energy density of
the $Q$ balls dominates the Universe at the time when they decay.

As a concrete example, let us consider no-scale SUGRA. During
inflation, the Heisenberg symmetry then guarantees the vanishing of
the tree-level Hubble-induced mass. Let us further assume that
initially $\phi_* \sim \phi_{osc}$ so that the perturbations generated
by curvaton are not damped. $K$ can be computed from the
renormalization group equations (RGEs), which to one loop has the
form
\begin{equation}
  \frac{\partial m_i^2}{\partial t} = \sum_g a_{ig}m_g^2 - \sum_a
  h_a^2\left(\sum_j b_{ij} m_j^2 + A_a^2\right)\,, \label{rgroup}
\end{equation}
where $a_{ig}$ and $b_{ij}$ are constants, $m_g$ are the gaugino masses,
$A_j$ the $A$-terms, $h_a$ the Yukawa couplings, $t=\rm{ln}(M_X/\mu)$
with $M_X$ the GUT scale, and $m_i$ are the masses of the scalar
partners. All the soft SUSY breaking scalar masses vanish at
tree-level at the GUT scale except the common gaugino mass $m_{1/2}$.

The full renormalization group equations are given in
\cite{Inoue:1982pi}. We neglect all the other Yukawa couplings except
the third generation. We assume that the top, bottom and tau Yukawas
unify at the GUT scale and normalize the unified coupling through the
top quark mass by
\begin{equation}
  m_{top}(m_{top}) = \frac{1}{\sqrt{2}}
  \,h_{top}(m_{top})\,v\,\sin\beta\,,
  \label{topmass}
\end{equation}
where $m_{top}=174.3$ GeV \cite{Hagiwara:fs}, $v=246$ GeV and
$0<\beta<\pi/2$ is a free parameter constrained by LEP as
$\tan\beta>2.4$ \cite{unknown:2001xx}. We find that Yukawa coupling
unification does not produce the correct top-Yukawa coupling given by
Eq. (\ref{topmass}) unless $\tan\beta\gtrsim 2.9$. The unification is
actually supported by $\tan\beta\sim 40-50$ \cite{Hagiwara:fs}, so
that our calculation clearly covers the relevant range. In
\cite{Enqvist:2000gq} $\tan\beta=1$ and only the top-Yukawa was
taken into account, which is applicable for small $\tan\beta$.

The mass of the flat direction scalar $\phi$ is the sum of the masses
of squark and slepton fields $\phi_i$ constituting the flat direction,
$m_{\phi}^2 = \sum_i |p_i|^2m_i^2$, where $p_i$ is the projection of
$\phi$ along $\phi_i$, and $\sum_i |p_i|^2=1$. The parameter $K$ is
then given simply by \cite{Enqvist:2000gq}
\begin{equation}
    K = -\left. \frac{1}{2m_{\phi}^2}\, \frac{\partial m_{\phi}^2}
      {\partial t} \right|_{t=\log(M_X/\mu)}\,.
\end{equation}

\begin{figure*}
\vspace*{6.5cm}
\includegraphics{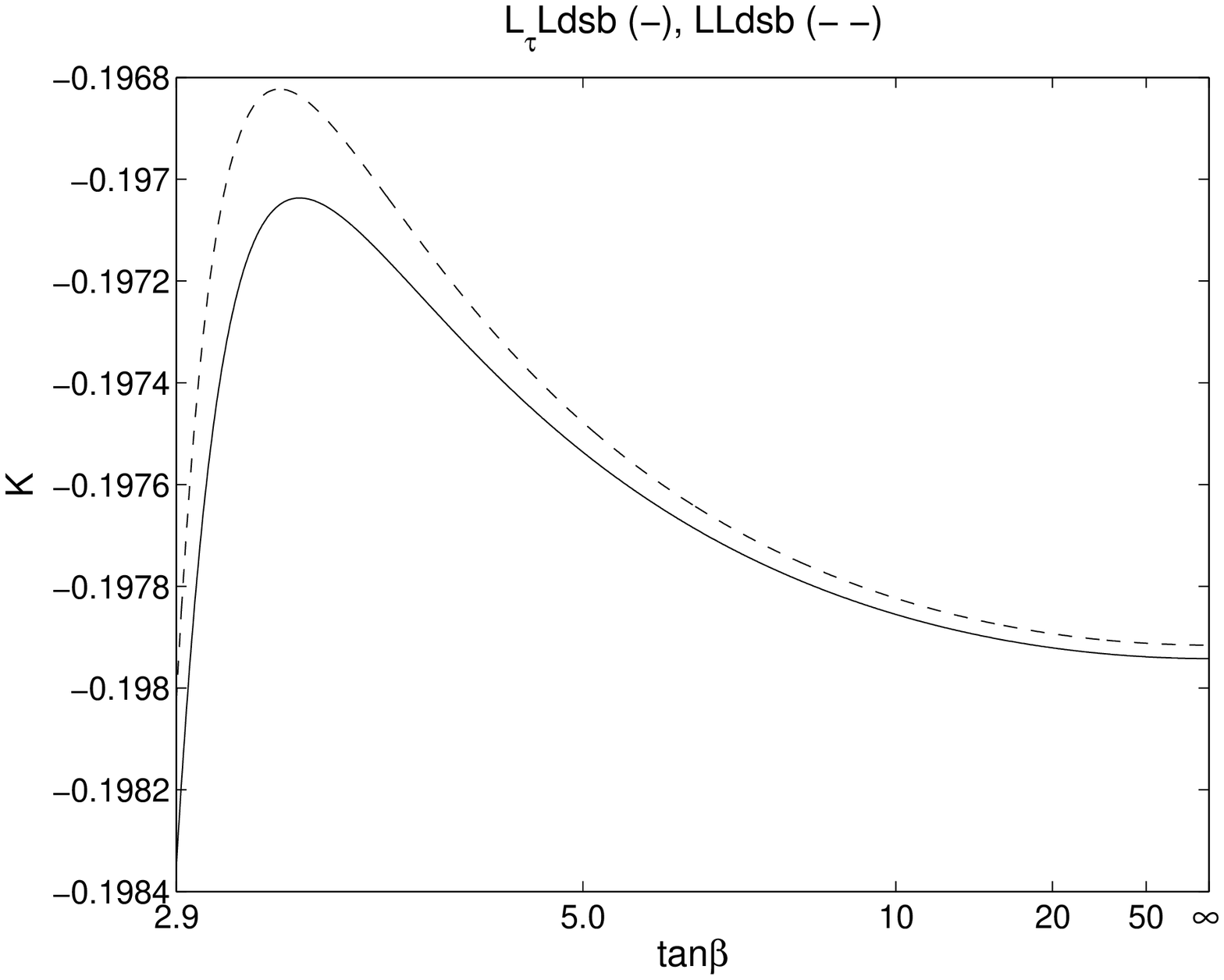}
\includegraphics{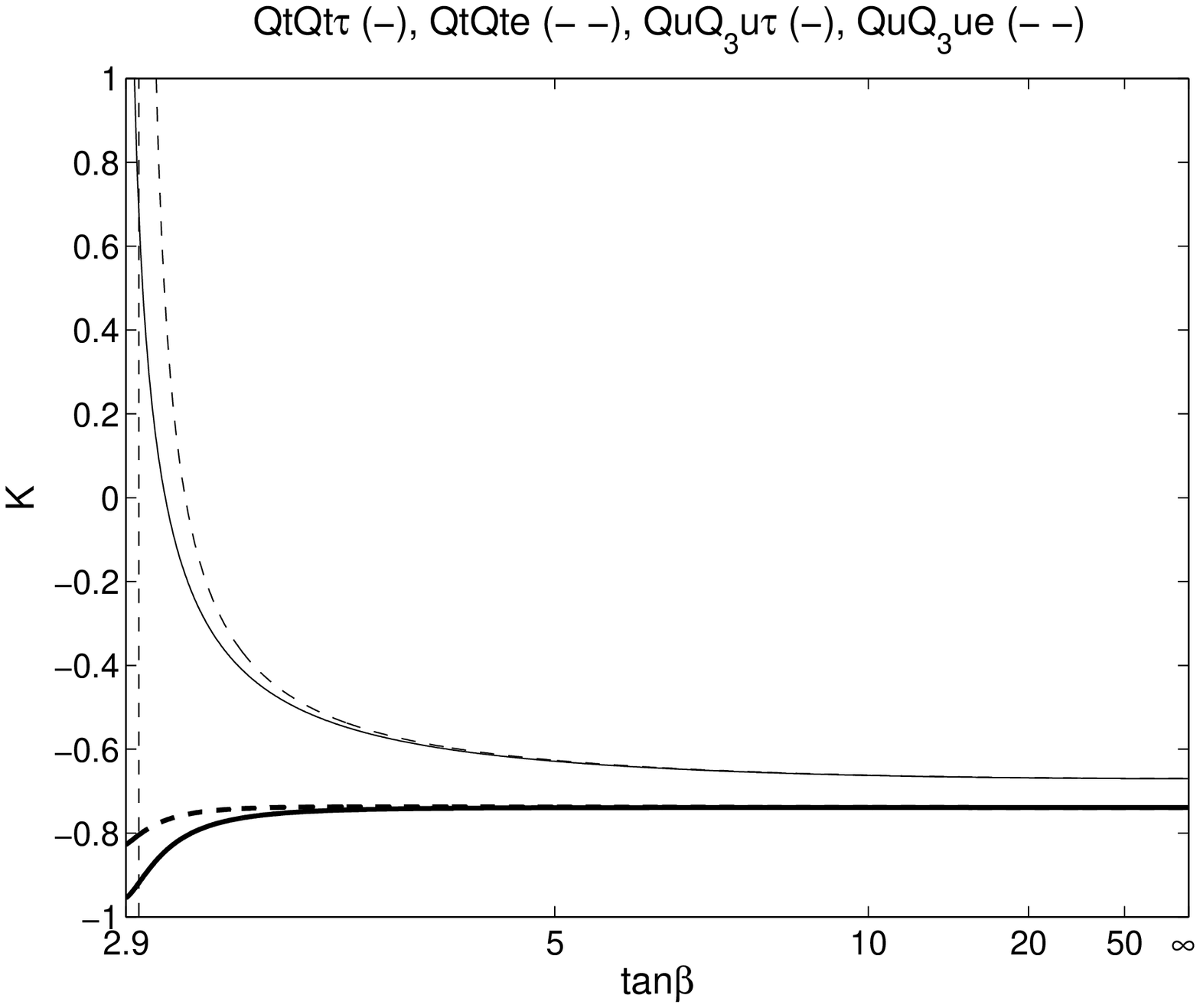}
\caption{\small $K$ vs. $\tan\beta$. On the left $LLdsb$ flat
direction with a choice of stau in the flat direction (solid line) and
no stau in the flat direction (dashed line). On the right $QuQue$ flat
direction with $u$ from 3rd generation (thick lines) and $Q$ from 3rd
generation (thin lines), $e$ from 3rd generation (solid line) and $e$
from 1st or 2nd generation (dashed line).}
\label{kfig}
\end{figure*}

To compute $K$, we have to choose the scale $\mu$. The appropriate
scale is given by the value of the flat direction field when it begins
to oscillate so that $\mu\sim\phi_{osc}$, see Eq. (\ref{osc}).
We have calculated $K$ for two flat directions: $n=7$ direction
$LLddd$ (lifted by $H_uLLLddd$) and $n=9$ direction $QuQue$ (lifted by
$QuQuQuH_dee$)\footnote{
We denote $LLddd$ by $LLdsb$ since all the family indices in
$ddd$ have to be different.}.
We find that $K$ is generically negative. In Fig.~\ref{kfig} we show
the coefficient $K$ plotted against the parameter $\tan\beta$ for the
two flat directions with different mixtures of stop, sbottom and
stau. In the $QuQue$ direction there are values of $\tan\beta$ where
$K$ is positive. This is due to the fact that for $n=9$ the
scale of oscillations, $\phi_{osc}\sim 1.5\times 10^{16}~\rm{GeV}$, is
very close to the GUT scale $M_X\sim 3\times 10^{16}~\rm{GeV}$. For
low $\tan\beta$ the Yukawa coupling at the GUT scale is large and
dominates over the gaugino terms in the renormalization group
equation, driving $\partial m_i^2/\partial t$ negative and making $K$
positive. When the renormalization group equations are run further,
the Yukawa couplings become smaller and thus $K$ becomes
negative. This is why the $LLdsb$ direction has a negative $K$ for all
$\tan\beta$.

Thus in most cases $K<0$ in no-scale SUGRA, so that the flat direction
condensates will not decay but instead fragment into lumps which
eventually form $Q$ balls. In general, these are long-lived and hence
give rise to a reheat temperature which is low (See, e.g.,
\cite{Enqvist:2003gh}). If R-parity is conserved, decaying $Q$ balls
will produce LSPs but with a low reheat temperature, their density
might come out to be too high \cite{Enqvist:1998en,Fujii:2002kr}. The
fragmentation of the flat direction condensate is yet another
complication for the MSSM curvaton scenario, which we do not attempt
to analyze systematically here. The sign of $K$ depends on the running
of the RGEs and hence on the initial conditions for the soft SUSY
breaking parameters, for which there is no generic form in the class
of SUGRA models with a Heisenberg symmetry.

\section{Conclusion}
To conclude, an MSSM flat direction curvaton appears to be very much
constrained, although not completely ruled out.  First of all, we have
argued that the constraints depend on the inflaton sector. If the
inflaton reheats the Universe with MSSM degrees of freedom, the finite
temperature effects on both the effective potential of the flat
direction and its decay (or evaporation) process are crucial. As
pointed out in Sect.II, the energy density of the flat direction
dominated by the thermal mass term cannot overcome the radiation
density, so that the curvaton will never dominate the Universe. This
seems to exclude flat direction curvatons in the presence of MSSM
radiation.

However, if the inflaton is completely in the hidden sector, there
will be no thermal corrections to the flat direction. In this case,
the curvaton has to provide both the adiabatic density perturbations
as well as dark and baryonic matter. The general requirements for this
scenario is the energy density dominance by the flat direction
condensate at the time of its decay, while during inflation its
contribution must be negligible. One ingredient is that during
inflation the Hubble-induced mass term should be negligible, a
situation that occurs in SUGRA models with a Heisenberg symmetry.

At large amplitudes the effective potential of the flat direction is
dominated by nonrenormalizable terms, and hence it is important to
follow the dynamics of both the homogeneous and fluctuation modes
during and after inflation. We have found that there is considerable
damping of the fluctuations, and in general it is hard to obtain a
successful curvaton scenario. Within one particular example of the
Heisenberg symmetry, the no-scale model, we have also shown that
typically the curvaton may fragment and form $Q$ balls rather than
decay directly, which will further complicate matters. Indeed, it is
not quite obvious whether $Q$ balls would be a help or a hindrance.

Damping of the fluctuations may however be avoided for a class of
initial values for the condensate field after inflation. Perhaps the
most promising candidate for the hidden inflation MSSM curvaton would
be the $n=9$ $QuQue$ 3rd generation direction with an initial
amplitude $\phi_* \sim 10^{-2} M_p$, based on a SUGRA model such that
there is no $Q$-ball formation at least in some parts of the parameter
space. In such a case one recovers the hot Universe at the
temperature $\sim 10^5$~GeV, which is high enough for
baryogenesis to occur during the electroweak phase transition
\cite{Kuzmin:1985mm}. Moreover, the reheat temperature is sufficiently
low in order not to create thermal or non-thermal gravitinos
\cite{ellis84}. More studies are nevertheless needed to settle the
open issues in more detail.

\section*{Acknowledgments}

S.K. is grateful to M. Kawasaki for useful discussions.
This work has been partially supported by the Academy of Finland
grant 51433.



\begin{thebibliography}{99}

\bibitem{Dine:1995kz}
M.~Dine, L.~Randall and S.~Thomas,
Nucl.\ Phys.\ B {\bf 458}, 291 (1996).


\bibitem{Gherghetta:1995dv}
T.~Gherghetta, C.~F.~Kolda and S.~P.~Martin,
Nucl.\ Phys.\ B {\bf 468}, 37 (1996).

\bibitem{Affleck:1984fy}
I.~Affleck and M.~Dine,
Nucl.\ Phys.\ B {\bf 249}, 361 (1985).


\bibitem{Dine:1995uk}
M.~Dine, L.~Randall and S.~Thomas,
Phys.\ Rev.\ Lett.\  {\bf 75}, 398 (1995).

\bibitem{Kusenko:1997si}
A.~Kusenko and M.~E.~Shaposhnikov,
Phys.\ Lett.\ B {\bf 418}, 46 (1998).

\bibitem{Enqvist:1998en}
K.~Enqvist and J.~McDonald,
Nucl.\ Phys.\ B {\bf 538}, 321 (1999).

\bibitem{Enqvist:1997si}
K.~Enqvist and J.~McDonald,
Phys.\ Lett.\ B {\bf 425}, 309 (1998).

\bibitem{Enqvist:1998pf}
K.~Enqvist and J.~McDonald,
Phys.\ Rev.\ Lett.\  {\bf 83}, 2510 (1999).

\bibitem{Enqvist:1999mv}
K.~Enqvist and J.~McDonald,
Nucl.\ Phys.\ B {\bf 570}, 407 (2000).

\bibitem{Kawasaki:2001in}
M.~Kawasaki and F.~Takahashi,
Phys.\ Lett.\ B {\bf 516}, 388 (2001).


\bibitem{Enqvist:2003gh}
K.~Enqvist and A.~Mazumdar,
arXiv:hep-ph/0209244.

\bibitem{Enqvist:2002rf}
K.~Enqvist, S.~Kasuya and A.~Mazumdar,
Phys. \ Rev. \ Lett. {\bf 90}, 091302 (2003).

\bibitem{Enqvist:2001zp}
K.~Enqvist and M.~S.~Sloth,
Nucl.\ Phys.\ B {\bf 626}, 395 (2002).

\bibitem{Lyth:2001nq}
D.~H.~Lyth and D.~Wands,
Phys.\ Lett.\ B {\bf 524}, 5 (2002);
D.~H.~Lyth, C.~Ungarelli and D.~Wands,
Phys.\ Rev.\ D {\bf 67}, 023503 (2003).




\bibitem{Others}
T.~Moroi and T.~Takahashi,
Phys.\ Lett.\ B {\bf 522}, 215 (2001);
Phys.\ Rev.\ D {\bf 66}, 063501 (2002);
N.~Bartolo and A.~R.~Liddle,
Phys.\ Rev.\ D {\bf 65}, 121301 (2002);
M.~S.~Sloth,
arXiv:hep-ph/0208241;
K.~Dimopoulos and D.~H.~Lyth,
arXiv:hep-ph/0209180;
M.~Bastero-Gil, V.~Di Clemente and S.~F.~King,
arXiv:hep-ph/0211011;
T.~Moroi and H.~Murayama,
Phys.\ Lett.\ B {\bf 553}, 126 (2003);
J.~McDonald, arXive: hep-ph/0302222.

\bibitem{Kolda:1998kc}
C.~F.~Kolda and J.~March-Russell,
Phys.\ Rev.\ D {\bf 60}, 023504 (1999).

\bibitem{Gaillard:1995az}
M.~K.~Gaillard, H.~Murayama and K.~A.~Olive,
Phys.\ Lett.\ B {\bf 355}, 71 (1995).


\bibitem{Lahanas:1986uc}
For a review, see A.~B.~Lahanas and D.~V.~Nanopoulos,
Phys.\ Rept.\  {\bf 145}, 1 (1987).



\bibitem{Postma:2002et}
M.~Postma,
arXiv:hep-ph/0212005.


\bibitem{Bennett:2003bz}
C.~L.~Bennett {\it et al.},
arXiv:astro-ph/0302207.

\bibitem{Kasuya:1999wu}
S.~Kasuya and M.~Kawasaki,
Phys.\ Rev.\ D {\bf 61}, 041301 (2000);
Phys.\ Rev.\ D {\bf 62}, 023512 (2000);
S.~Kasuya,
in: {\it Proceedings of the 7th International Symposium on Particles,
Strings and Cosmology (PASCOS-99)}, K. Cheung et al., 301 (1999).


\bibitem{Inoue:1982pi}
K.~Inoue, A.~Kakuto, H.~Komatsu and S.~Takeshita,
Prog.\ Theor.\ Phys.\  {\bf 68}, 927 (1982)
[Erratum-ibid.\  {\bf 70}, 330 (1983)];
Prog.\ Theor.\ Phys.\  {\bf 71}, 413 (1984).

\bibitem{Hagiwara:fs}
K.~Hagiwara {\it et al.}  [Particle Data Group Collaboration],
Phys.\ Rev.\ D {\bf 66}, 010001 (2002).


\bibitem{unknown:2001xx}
LEP Higgs Working Group Collaboration,
arXiv:hep-ex/0107030.




\bibitem{Enqvist:2000gq}
K.~Enqvist, A.~Jokinen and J.~McDonald,
Phys.\ Lett.\ B {\bf 483}, 191 (2000).


\bibitem{Fujii:2002kr}
M.~Fujii and K.~Hamaguchi,
Phys.\ Rev.\ D {\bf 66}, 083501 (2002).



\bibitem{Kuzmin:1985mm}
V.~A.~Kuzmin, V.~A.~Rubakov and M.~E.~Shaposhnikov,
Phys.\ Lett.\ B {\bf 155}, 36 (1985).


\bibitem{ellis84}
J. Ellis, J E Kim, and D. V. Nanopoulos,
Phys. Lett. B {\bf 145}, 181 (1984);
J Ellis, D V Nanopoulos, K A Olive and S-J Rey,
Astropart Phys. {\bf 4}, 371 (1996);
A. L. Maroto, and A. Mazumdar,
Phys. Rev. Lett. {\bf 84}, 1655 (2000);
R. Kallosh, L. Kofman, A. D. Linde and A. Van Proeyen,
Phys. Rev. D {\bf 61}, 103503 (2000);
R. Allahverdi, M. Bastero-Gil and A. Mazumdar,
Phys. Rev. D {\bf 64}, 023516 (2001).



\end{thebibliography}
\end{document}